# SQR: Balancing Speed, Quality and Risk in Online Experiments

Ya Xu
LinkedIn Corporation
950 W Maude Ave
Sunnyvale, CA 94085
yaxu@linkedin.com

Weitao Duan
LinkedIn Corporation
950 W Maude Ave
Sunnyvale, CA 94085
wduan@linkedin.com

Shaochen Huang
LinkedIn Corporation
950 W Maude Ave
Sunnyvale, CA 94085
sahuang@linkedin.com

## ABSTRACT

Controlled experimentation, also called A/B testing, is widely adopted to accelerate product innovations in the online world. However, how fast we innovate can be limited by how we run experiments. Most experiments go through a "ramp up" process where we gradually increase the traffic to the new treatment to 100%. We have seen huge inefficiency and risk in how experiments are ramped, and it is getting in the way of innovation. This can go both ways: we ramp too slowly and much time and resource is wasted; or we ramp too fast and suboptimal decisions are made. In this paper, we build up a ramping framework that can effectively balance among Speed, Quality and Risk (SQR). We start out by identifying the top common mistakes experimenters make, and then introduce the four SQR principles corresponding to the four ramp phases of an experiment. To truly scale SQR to all experiments, we develop a statistical algorithm that is embedded into the process of running every experiment to automatically recommend ramp decisions. Finally, to complete the whole picture, we briefly cover the auto-ramp engineering infrastructure that can collect inputs and execute on the recommendations timely and reliably.

## 1 INTRODUCTION

There is no doubt that experimentation, or A/B testing, has become a driving force of innovation in the online world. It is not just the established players who have bought into the value of experimentation, as shared in several past papers from Microsoft, Google and LinkedIn [1,2,3]. Startups and smaller websites have also invested in building out their experimentation program as a necessity of growth [4].

One primary reason that companies have relied on A/B testing is that it accelerates product innovation. It is as if you have a "crystal ball" that can tell us how our users will react and how the business metrics will change if a new feature is rolled out. We can learn quickly, and ultimately, build better products faster. However, how fast we innovate can also be limited by how we run experiments. This becomes more apparent the more experiments we run. At companies like LinkedIn, where experimentation is truly embraced as a default step in every product change, this issue can be even more magnified.

At LinkedIn, we run over 5,000 experiments a year. Every experiment goes through a "ramp up" process. A new feature usually starts out by ramping to a small percentage of users, waits to see if metrics are good, and then repeats by ramping up to a higher percentage, until finally it reaches 100%. Such "ramp up" process is a standard practice across the industry to control unknown risks associated with any new feature launches. However, many times the way we ramp slows us down. This can go both ways: we ramp too slowly and much time and resource is wasted; or we ramp too fast and suboptimal decisions are made. On average, an experiment at LinkedIn takes four ramps to reach 100%, where each ramp takes about six days – that's almost a month from start to finish! In addition, experimenters tend to treat each of the incremental ramps the same. For example, on average, we spend 6 days waiting on a 5% ramp, while 6.5 days on a 50% ramp. So having 4 ramps for one experiment is almost equivalent to running 4 separate experiments sequentially!

The message behind these numbers is loud and clear: while we can democratize experimentation with a fully self-served platform [3], we need principles to guide us on how we should ramp. Running slower doesn't mean we are safer. Taking longer to finish an experiment doesn't mean we are more cautious regarding negative user impact. It is important to point out that while there is a need for "speed", the principles should not be driven by "ramping as quickly as possible", but by "ramping with the right balance of Speed, Quality and Risk". The core of our paper is to answer the following question: **how can we go fast while controlling risk and improving decision quality**?

At first, the answer may seem to lie with "power analysis." Power analysis is widely used to calculate the minimum sample size required to detect a given effect size [5]. In the world of online A/B testing, where samples trickle into experiment continuously, this usually translates to deciding on both percentage of traffic ramped to the treatment (ramp percentage) and also the duration for the experiment. However, power analysis fails to serve our needs. (1) It can only be performed for one metric at a time. With hundreds of metrics we monitor closely for each experiment [3], summary from power analysis of individual metrics becomes uninterpretable. (2) What is considered as "enough power" can be different depending on the stage of the ramp process. In general, we can tolerate a lower power (i.e. higher Type II error) during earlier ramps. We will discuss more formally about this in Section 4. (3) Power analysis alone is not actionable. For most experimenters, power is a hard concept to digest, and is even harder to take action upon. If my experiment does not have enough power, what should I do?

After several attempts at making our ramp process more scientific and efficient, including an introduction and deprecation of power analysis on our experimentation platform, we have arrived at a solution that has been hugely successful at LinkedIn, called SQR, for **S**peed, **Q**uality and **R**isk. The SQR framework divides up the whole ramp process into four phases, each with a primary goal to achieve. The first phase is mainly for risk mitigation, so the SQR framework focuses on trading off between speed and risk. The second phase is for making precise measurement, so the focus is on trading off between speed and quality. The last two phases are optional and are used to address additional operational concerns (third phase) and long-term impact (forth phase). The first two phases are common to all experiments and are usually the only two phases an experiment needs. Most of our effort in this paper is dedicated to addressing how to ramp through these two phases.

In most cases, each step in ramping an experiment is done by engineers and product managers. The principles that SQR offers need to be as simple and straightforward as possible to be accessible to all experimenters. While we can gain a lot of efficiency simply by following the principles manually, we need to embed the principles as part of the ramp process in an automatic fashion to fully scale to all experiments. This requires a whole suite of solutions, including both a statistical algorithm that can recommend ramp decisions and an engineering infrastructure that can collect input and execute on the recommendations reliably. We will discuss the algorithm in depth in Section 4, as it is probably most interesting to the KDD community, while briefly going over the auto-ramp infrastructure in Section 5 to complete the whole picture.

Here is a summary of our contributions from this paper:
- As far as we know, we are the first to conceptualize the need of balancing among speed, quality and risk in online experiments, and also the first to study it extensively.
- We offer simple and straightforward principles that any practitioner can follow to effectively balance speed, quality and risk for their experiments.
- We develop rigorous statistical procedures that algorithmically recommend next steps in the experiment ramping process to achieve SQR.
- We share the engineering infrastructure that allows us to automatically ramp experiments in a reliable and timely fashion.

The paper is organized as follows. Section 2 starts with a review of the existing literature in both areas of A/B testing and sequential experiments. Section 3 develops the SQR ramping framework, describing common mistakes that people make and offering easy-to-follow principles to guide experimenters. Section 4 focuses on the Ramp Recommender algorithm that sets the foundation of automating the ramping process. Section 5 briefly goes over the auto-ramp infrastructure. Section 6 concludes with future work.

## 2 LITERATURE REVIEW

In this section we review the evolution of controlled experiment theories and applications, especially in the field of sequential testing. The foundation of experimentation theory was first introduced by Sir Ronald A. Fisher in the 1920s with a focus on agricultural activities [6]. Since then, this subject has been studies by many researchers in papers and textbooks [7,8]. Controlled experiment has gained its popularity beyond the original agricultural and manufacturing industries, and has been widely adopted across companies including commercial and tech sectors. In particular, past papers have discussed experimentation at scale in Microsoft Bing, Google, Facebook and LinkedIn [1,2,9,3], sharing success stories, best practices and pitfalls.

Despite the benefits of running an experiment, the associated cost and logistical constraints cannot be overlooked. Being able to stop an experiment early and iterating faster has been a focus for many researchers. Abraham Wald first formulated and studied the sequential testing problem [10]. In the past seventy years, researchers are drawn to this field and contributed to the theoretical foundation [11,12]. Sequential testing methodology has been widely adopted in clinical trials [13]. Recently, it has gained its popularity in educational/psychological testing [14]. Although many companies adopt online A/B testing, as far as we know, among large scale online A/B testing platforms, Optimizely is the first to apply sequential testing methodology in experiment evaluation process [15]. However, the application there focuses on sequential monitoring of single stage A/B experiment with no practical considerations on balancing speed, quality and risk. In this paper, we utilize sequential testing and tailor it to fit the goals in different ramping stages.

In this paper, we leverage extensively the basic terminology and knowledge of the field of A/B testing, such as hypothesis testing, t-test, power etc. Readers who are not familiar with these topics are encouraged to read the survey and practical guide by Kohavi et al. [5].

## 3 SQR FRAMEWORK

As we discussed in Section 1, an online experiment usually goes through a "ramp up" process, where it starts with a small percent of traffic for the new treatment and then gradually increases the percentage to 100%. On a fully self-served platform, experimenters are free to choose whichever incremental ramps they want, any fractions between 0% and 100%. As a matter of fact, people chose over 300 unique ramp sequences for their experiments at LinkedIn in year 2015, with an average of four ramps per experiment. We may expect some diversity in the ramp process due to the diversity of our experiments, but 300 is an astonishingly big number. People are also spending a lot of time at each ramp. Figure 1 plots the distribution of time-spent on ramps at a given ramp%. While 1% ramps tend to go a little faster, all the other common incremental ramps tend to take similar amount of time to finish (about 6 days on average). All these numbers are indications that leaving people to decide how they ramp can be extremely inefficient and that it is getting in the way of innovation. We need principles, and better yet, automatic algorithms embedded into the ramp process of every experiment. Simply put, if we can cut the experiment length by half with risk controlled and decision quality improved, we can double the amount of new things we try.

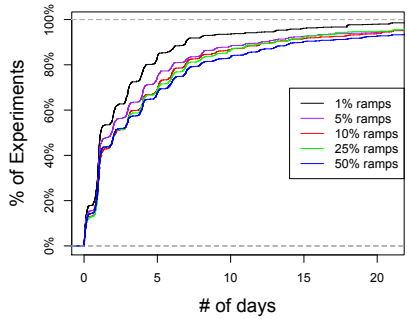

Figure 1: Cumulative distribution of ramp duration, by ramp%.

Our goal in this section is to come up with a framework to ramp that can achieve the balance between speed, quality and risk. The framework should be simple yet comprehensive. We want all our experimenters to be able to follow and apply to their own experiments, regardless whether they are product managers, engineers or analysts. Let's start with understanding why people ramp the way they do.

### 3.1 Three Mistakes Most People Make

In this section, we cover the top three misconceptions that people have when it comes to deciding how to ramp their experiments. These mistakes may seem simple, but they are so prevalent among experimenters that during one of our experimentation training sessions to engineers and product managers, almost all hands were up when asked who has made at least one of these mistakes. Understanding these mistakes is our starting point of creating the SQR framework.

**Mistake #1: Let's keep it running to get statistical significance.**

At LinkedIn, we have been very successful at getting the whole company to care for "statistical significance." People understand that they shouldn't read too much into the results unless they are "significant." However, almost all people assume if they run their experiments long enough, they will eventually get statistically significant results. Of course, this assumption could be true simply because of a higher false positive rate due to multiple testing over a longer period of time, a mistake that Optimizely founders made in their popular book [16]. Besides that, this assumption is not entirely unfounded for online experiments. As we accrue users over time, we do tend to get a bigger sample size as we run the experiment longer. So given the same effect size, we are more likely to get statistically significant results. However, the tradeoff between running longer vs. ramping up is rarely considered.

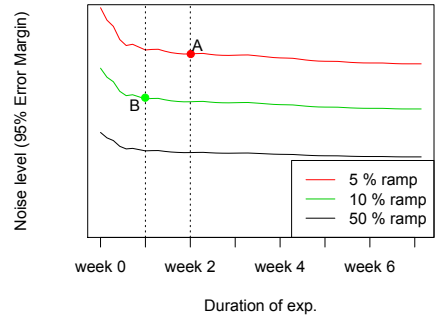

Figure 2: Tradeoff between ramping up (B) vs. running longer (A).

We did a poll in the company asking people "Which of the following is more likely to result in statistically significant results: running experiment at (A) 5% for 2 week, or (B) 10% for 1 week?" Most people chose (A). The fact is, for almost all our metrics, the benefit of running longer drastically diminishes after 1 week, which means ramping up to a higher traffic is almost always a better option than waiting longer. As shown in Figure 2, the confidence interval for the percent delta (similar to Coefficient of Variation) is much narrower for a 10% ramp after 1 week (option B) than for a 5% ramp after 2 weeks (option A).

**Mistake #2: The cost is lower if I keep the experiment at a smaller ramp (even for a longer time).**

It is correct that the cost tends to be smaller with a lower ramp. However, this is not always the case, especially when we keep the experiment at the low ramp for a long time to wait for significance, without a decision to either ramp down or ramp up. As we have just discussed, there is little gain in statistical power, yet it is costly to keep an experiment alive for a long time.

1) **Opportunity cost**. There is certainly time and effort from experimenters to keep an experiment alive, which directly translates to fewer and slower innovations.
2) **Platform cost**. With a highly automated experimentation platform, the incremental cost-to-experiment is small. However, at the scale of thousands of experiments a year, the cost does add up.
3) **Business cost**. There can be a combination of loss on both user engagement and revenue. Putting users in a bad experience for a long time is more likely to cause permanent user churn. The cumulative loss can also be high over time. Assume that an experiment has a real negative impact of -1% on revenue. Such negative impact is too small to be detected at 10% ramp, even if we keep the experiment running for two weeks. The revenue lost during the two-week 10% ramp is not only in vain, but also turns out to be more than the loss if running the experiment at 50% for two days, which does have sufficient power to detect a -1% impact.

**Mistake #3: We have enough users at the 10% ramp. Let's ramp to 100%.**

It is hard for people to believe that at the scale of LinkedIn's traffic, when we have millions of users visiting our site everyday, we still need more users for our experiments. The reason is two-fold. First, many of our experiments are effective for only a small subset of our users. As a matter of fact, 60% of our experiments are triggered for less than 10% of active users. So the real sample size is much smaller than millions. Second, many of our metrics are extremely volatile, especially revenue related metrics. These metrics need a higher volume for the normality assumption to be plausible according to Central Limit Theorem [17]. Their variance is usually too high for conclusions to be drawn at a lower ramp. Because of these reasons, there is always a need for better resolution with more users, so that experiments with real impact can be identified (even if the impact is unintended). We have seen similar discussions from Google [2] and Microsoft [18] with their scale of traffic too.

## 3.2 SQR Principles

Now that we have understood what has been preventing the experiments from being efficiently ramped, we are ready to introduce the SQR principles that can guide experimenters to properly balance Speed, Quality and Risk. Let's first answer why we do an A/B experiment at all. The following three reasons cover almost all scenarios:

1) **To measure**: measure the impact and ROI (Return-On-Investment) to decide whether to launch the new treatment to 100%.
2) **To reduce risk**: reduce the damage and cost to both users and business during the experiment in case of negative impact.
3) **To learn**: learn about users to help us innovate better. This can be a general purpose as part of running any experiments, or some specific experiments designed specifically for learning purpose, e.g. site speed degradation tests [19].

To see how each of the goals maps onto the ramp process, let's first introduce the concept of **Maximum Power Ramp** (MPR). MPR is the ramp that gives the most statistical power to detect differences between treatment and control. It has the smallest variance, and hence the delta measured has the best precision. If the experiment has the entire 100% traffic with only one treatment, the variance in the two-sample t-test is proportional to $1/q(1-q)$ where $q$ is the treatment traffic percentage. Hence, the MPR in this case is 50%. If there is only 20% traffic available to experiment between one treatment and one control, the MPR is 10%, and so on. For simplicity, unless specified otherwise, we assume that MPR is 50% because it is the most common case for LinkedIn given that most of our experiments are orthogonal.

If the only reason to run an A/B test is "to measure," we should always ramp at the Maximum Power Ramp. Statistically, this gives us the fastest and the most precise measurement. But of course we cannot start at 50% ramp, at least not in most cases - what if something goes wrong? That's why we usually start at a small ramp, with the goal to contain impact and mitigate potential risk.

In certain situations, we also need intermediate ramps between MPR and 100%. For example, it is often that for operation reasons we need to make sure the new services or endpoints are able to handle the increasing traffic load. In these cases, we need to stop at extra ramps (e.g. 75%) to make sure the service metrics are stable before fully ramping to 100%. Another common example is to learn. While learning should be part of every ramp, we sometimes conduct a long-term holdout ramp primarily for learning purpose. This is a ramp where only a small fraction of users are kept out of the new experience (e.g. 5%) for a long time, usually over a month. The goal is to learn whether the impact measured during MPR is sustainable in the long run.

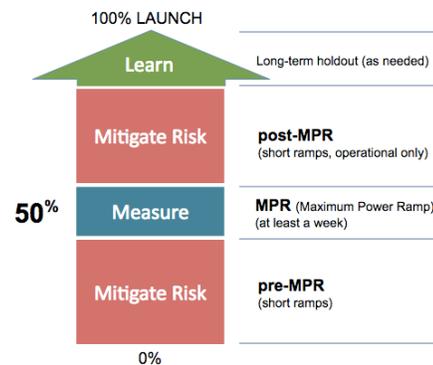

**Figure 3:** SQR Ramp Process

Putting all the pieces together, Figure 3 shows the four phases a ramp process goes through and how they each map to the three goals we have: (1) the pre-MPR ramps primarily for mitigating risk; (2) the MPR primarily for measuring impact; (3) the optional post-MPR ramps used for mitigating operational risks; (4) the optional long-term holdout ramps used for learning. We are now ready to present the SQR principles.

**Principle #1: Ramp quickly to the Maximum Power Ramp (MPR), as soon as the risk is determined small.**

With this first principle, we are not asking teams to take more risk by going faster, but to be cognizant that earlier ramps are there for risk mitigation and not for decision-making. So as soon as we are comfortable with the risk, we should ramp further. The earlier ramps and the MPR should be treated vastly differently – we should not be "waiting" for statistical significance at any pre-MPR ramps. The most common mistake we have heard people saying is "nothing is significant yet, let's keep the ramp at 5% for longer".

As core to the first principle, we need to decide what is risk. We will introduce more formally the definition of risk and how to quantify it in Section 4. In general, risk assessment should include a combination of prior knowledge and risk learnt from data. For example, an experiment that changes the font color does not expect to impact either product or operational metrics. So risk is initially assessed as "low." After one day at 5%, we may notice some unexpected impact on metrics. At that time, the risk is reassessed to be higher according to the data and we may not be able to ramp further. Another important factor to include in risk assessment is trigger rate. By and large, trigger rate refers to the percentage of traffic volume that is affected by the experiment. As we mentioned earlier when discussing Mistake #3, not all experiments impact all users. For example, some experiments are only triggered if a user is on an older mobile app version. For such experiments with low trigger rate, the total impact tends to be smaller, and hence the risk should be lower as well.

**Principle #2: Spend enough "waiting" time at MPR.**

Because MPR is the ramp dedicated to measure the impact of the experiment, it is crucial that we not only have the most precision but also capture any other time-dependent factors. For example, an experiment that runs for only one day will have results that are biased towards heavy users. Another example is burn-in effect. Many of our new features that involve drastic UI changes have impact (either positive or negative) that dies down over time. Because there is usually little gain on precision after one week and because we want to capture at least one full week, we advise all our teams to keep their experiment at MPR for at least a week, and longer if burn-in effect is present.

**Principle #3: Conduct quick post-MPR ramps if there are operational concerns.**

By the time an experiment is passed the MPR phase, there should be no additional concerns regarding end user impact. In most cases, operational concerns should also be resolved in earlier ramps. There are some cases where we worry about increasing traffic load to some engineering infrastructure that warrants incremental ramps before going to 100%. These ramps should only take a day or less, usually covering peak traffic period, with close monitoring.

**Principle #4: Conduct optional long-term holdout ramps only if the learning objectives are clear.**

We have seen an increasing popularity of long-term holdout ramps. In most cases, the experiment is left running for a long time, and eventually terminated without any learning. The decision to do a long-term holdout should be made with a clear objective. In general, we should only do a long-term holdout if we have a strong belief that users may interact with the new feature differently over time. It is not a default last step in the SQR ramping model, neither is it a panacea to any questions leftover from an earlier ramp.

We have identified a few scenarios where long-term holdout can be helpful. The most common scenario is when an experiment has a long-lasting burn-in effect. There are some known experiment areas where burn-in exists by design, such as the People You May Know recommendation [3]. Another example is when a feature takes time for users to discover. Even though the short-term observed impact is zero, we believe there is benefit in the long term. The last example is when the experiment shows big impact during MPR for topline business metrics. Because such impact is to be baked into financial forecasting, we need to know whether the effect sustains in the long run.

## 4 RAMP RECOMMENDER

To execute the SQR principles, we rely on an algorithm driven approach to automate the ramping process as much as possible. One core piece of the automation is the underlying algorithm that provides the statistical foundation. This section is dedicated to introduce the ramp recommendation algorithm. The engineering infrastructure that executes the recommendation will be covered in Section 5.

SQR framework suggests that we should ramp as quickly as possible to the Maximum Power Ramp, and spend enough time to measure experiment impact at MPR. Correspondingly, Ramp Recommender performs two tasks depending on the ramp phase: (1) Guide the ramps towards MPR. (2) Give signal to ramp up from MPR. We will cover each in Section 4.1 and 4.2 respectively. The last two phases of the ramp process, post-MPR and long-term holdout, are both optional with relatively simple ramp criteria, and hence are omitted from the discussions here.

### 4.1 Ramping towards MPR

This is the phase where we need to effectively trade off between speed and risk concerns. We would like to ramp as quickly as possible to MPR, but we need to make sure the risk is tolerable. We formulate the problem first by quantifying risk and then by developing a rigorous statistical procedure to test the risk. Intuitively, at every time point $t$, we utilize all the information we have up to that point, including both prior knowledge and information learnt from data, to answer the question whether the risk is small enough to ramp up, and if so, ramp to what percentage of traffic. We start assuming there is only one metric of interest, and then extend to cases with multiple metrics. Without loss of generality, we use "day" as the default time unit.

*4.1.1 Risk and Tolerable Risk.* We start with quantifying risk. Because we ultimately need to translate "risk" into a binary decision of ramping up or not, our definition includes both risk and risk tolerance. Intuitively, if the risk is below the risk tolerance, we can ramp up. Otherwise, we cannot. We define the risk of ramping to traffic percentage $q$ as

$$R(q) = |\delta| * g(r) * h(q)$$

where

$$\delta = \frac{treatment\ mean - control\ mean}{control\ mean}$$

captures relative impact on the triggered population, and

$$g(r) = \begin{cases} r, & r \geq r_0 \\ r_0, & r < r_0 \end{cases}$$

is trigger rate $r$ truncated at $r_0$, and

$$h(q) = \begin{cases} q, & q \geq q_0 \\ q_0, & q < q_0 \end{cases}$$

is the ramp percent $q$ truncated at $q_0$. Naturally, the risk is higher for a higher trigger rate experiment that has a bigger impact and is ramped to more users. The reason that we choose a truncated version of both trigger rate and ramp percent is because we do consider a really bad experiment (large $\delta$) too risky for our users even if it is only impacting a small set of users. Other reasonable, monotonically increasing functions can be used depending on the business consideration. Also note that we do not restrict risk to negative metrics impact only. We have seen many real examples where a positive move on a metric turns out to be bad [20].

We say that the risk of ramping to $q$ is *tolerable* if it is below the risk tolerance $\tau$, i.e.
$$R(q) \leq \tau,$$
where $\tau$ is set based on business requirement and is different for different metrics. It is usually decided by metrics owners through answering "As an organization, we do not want any experiment to hurt the overall health of metric X by more than $\tau$ for a day."

*4.1.2 Hypothesis Testing.* With risk defined, we are now ready to formulate the problem in terms of hypothesis testing. Let $Q = \{q_1, q_2, q_3, q_4, \dots\}$ be the set of possible ramps. For practical reasons, we want to restrict the cardinality of $Q$ to a reasonable number and use only the representative ramp percentages. For example, at LinkedIn we use $Q = \{1\%, 5\%, 10\%, 25\%, 50\%\}$.

The first ramp is determined based on the initial risk assessment by the experimenter. Naturally, the higher the risk is, the smaller the first ramp. With data from the initial ramp, we can then test to see whether risk is low enough for us to ramp further to the next ramp $q$. We can formulate this in the following hypothesis test. For any potential next ramp $q \in Q$, we have:
$$H_0^q: R(q) \leq \tau$$
$$H_1^q: R(q) > \tau \qquad (1)$$
Notice that the risk function is monotonically increasing in the ramp percent $q$. Therefore, for any $q_1 \leq q_2$, if $H_0^{q_2}$ is accepted then $H_0^{q_1}$ is also accepted. In other words, if we can safely ramp to $q_2$, we can safely ramp to a smaller ramp $q_1$ as well. The Ramp Recommender takes on a greedy approach that picks the maximum ramp among all feasible ramps. After the experiment is ramped to $q$, the hypothesis testing process repeats until reaching MPR.

Following the first SQR principle, we would like to ramp quickly and efficiently during the pre-MPR phase. Therefore, instead of using a fixed duration for each ramp, we want to evaluate the possibility of ramping up continuously as more data become available. With the large sample sizes that come with online experiments, t-test or z-test [5] can be the choice to determine which region $R(q)$ falls into. However, performing the above hypothesis testing everyday can easily inflate the type I/II error rates due to multiple testing [17]. For example, if we continuously test for risk, and ramp up as soon as risk is lower than the threshold, we have a higher chance of ramping up a risky feature. Sequential testing, however, allows continuously monitoring of risk $R(q)$ while restricting type I/II errors. We introduce sequential hypothesis testing techniques in the section below.

*4.1.3 Sequential Testing.* We use the Generalized Sequential Probability Ratio Test (GSPRT) [21]. At time $t$, the test statistic for $H_k^q$ is constructed as follows:
$$L_t(H_k^q) = \frac{\sup_{H_k^q} \pi_k f_k^t(\boldsymbol{X^t})}{\sum_{j=0}^{1} \sup_{H_j^q} \pi_j f_j^t(\boldsymbol{X^t})}, \qquad k = 0, 1 \qquad (2)$$

where $f_k^t$ is the likelihood function, $\boldsymbol{X^t} = (X_1^t, X_2^t, \dots)$ is the user-level metric value up to time $t$, and $\pi_k$ is the prior probability for hypothesis $H_k$. Note that it is important to consider prior risk assessment. In most cases, experimenters have an expectation whether their experiments may be impacting certain metrics or not. For example, many of our infrastructure experiments are not expected to move metrics at all, and hence the priors for such experiments should have $\pi_0 \gg \pi_1$.

Following GSPRT, the hypothesis $H_k^q$ is accepted if
$$L_t(H_k^q) > \frac{1}{1 + A_k}$$
with $A_k$ chosen to control the errors of accepting $H_k^q$ incorrectly. Note that since the posterior probabilities $L_t(H_0^q) + L_t(H_1^q) = 1$, we can choose $0 < A_k < 1$ to ensure that at most one hypothesis $H_k, (k = 1,2)$ is accepted [21].

Figure 4 below demonstrates the three regions that the test statistic $L_t(H_0^q)$ can fall in: the acceptance region, the waiting region and the rejection region. Note that we can construct an equivalent set of regions based on $L_t(H_1^q)$ with thresholds $A_0/(1 + A_0)$ and $1/(1 + A_1)$. If $L_t(H_0^q)$ falls into the rejection region, the experiment is considered too risky to be ramped to $q$. If $L_t(H_0^q)$ falls into the acceptance region, it is considered safe to ramp to $q$. If $L_t(H_0^q)$ falls in between, we do not have enough evidence to support either of the hypotheses, and hence we keep the experiment running at the current ramp to collect more data.

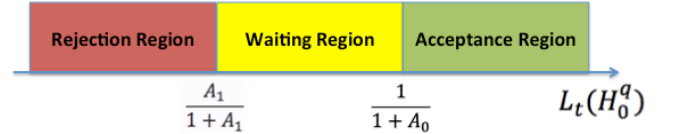

**Figure 4: Three regions the test statistic can fall into.**

The explicit form of $f_k^t$ is unknown and different for different metrics. When the sample sizes are large, based on multivariate Central Limit Theorem, the likelihood function of $\Delta$, the relative difference of the sample means, approaches normal [17]. The test statistic in Equation (2) becomes
$$L_t(H_k^q) = \frac{\sup_{H_k^q} \pi_k \exp\left(-\frac{(\Delta - \delta)^2}{s^2}\right)}{\sum_{j=0}^{1} \sup_{H_j^q} \pi_j \exp\left(-\frac{(\Delta - \delta)^2}{s^2}\right)}$$

where $s^2$ is the variance of $\Delta$ that can be estimated from the data and $\delta$ is the population parameter defined earlier that we are testing in the hypothesis Equations (1). We omit the time parameter $t$ in some notations here to keep it easier to read.

To see how we can choose $A_k$ to control the errors of accepting $H_k^q$ incorrectly, let $\alpha_0$ be the probability that $H_0$ is accepted if $H_1$ is true. Similarly, $\alpha_1$ is the probability of accepting $H_1$ if $H_0$ is true. Note that $\alpha_0$ is equivalent to the "usual" Type II error, while $\alpha_1$ corresponds to the "usual" Type I error. Intuitively, assuming $H_1$ is true, it is less likely to accept $H_0$ incorrectly with a smaller $A_0$ (and thus a bigger $1/(1 + A_0)$). It has been shown that the errors $\alpha_k$ can be bound by the choices of $A_k$, i.e. $\alpha_k \leq A_k$ [21].

To guide organizations on what bounds to use for such "Type I and II" errors, it is helpful to think about these errors as a tradeoff between speed and risk. When a Type I error is made, the algorithm says we shouldn't ramp up when we should, so we are going too slowly. When a Type II error is made, we ramp up when the risk is actually higher than the threshold we set, so we are taking too much risk. At LinkedIn, we have infrastructure in place to identify bad experiments quickly, so we are confortable with a higher Type II error and a lower Type I error in the pre-MPR phase:

$$A_0 = 0.2, A_1 = 0.01.$$

Putting it together, the algorithm follows the steps below. On any day $t$,

1) If $L_t(H_1^q) > 1/(1 + A_1)$ for every possible $q \in Q$, we accept $H_1^q$. The metric impact is deemed to be severely worse than the tolerable threshold. We cannot ramp the treatment further.
2) If for some $q \in Q$, $L_t(H_0^q) > 1/(1 + A_0)$, we can accept $H_0^q$. Ramp $q$ is deemed to be risk tolerable. As we discussed before, if multiple $q$'s are risk tolerable, we take the greedy approach and ramp to the largest $q$ acceptable.
3) Otherwise, there is not enough evidence supporting any of the hypotheses. Continuing running the experiment on day $t + 1$ and evaluate $L_{t+1}$.
4) If by day $t = 7$, none of the hypothesis can be accepted, we assume no effect is detected, and recommend ramping.

*4.1.4 Multiple Metrics.* We have so far assumed there is only one metric of interest. The reality is that for most experiments, there are over one hundred metrics that are closely watched. There are company-wide metrics that are the most important for all experiments. In addition to that, each experiment has its own success metrics. These metrics tends to capture more direct impact from the experiment and are likely to be local to a particular product area. As we mentioned in Section 4.1.3, different prior risk $\pi_k$ can be used for each metrics depending on how likely we expect them to be impacted by the experiment. From our experience, simplify the prior risk input to only two or three categories help make it actionable for everyone.

We can follow the testing procedure in Section 4.1.3 for every important metric, but we still need to combine the results and come up with a single ramp decision. Naturally, if *one* metric is truly impacted beyond our tolerance, we should not ramp up. On the other hand, we should ramp up only if we are confident that risk is small across *all* the important metrics. However, if we accept $H_1^q$ as long as it is accepted for one metric, we inflate the chance of accepting $H_1^q$ incorrectly (Type I error) due to multiple testing; Similarly, if we accept $H_0^q$ only if it is accepted for all metrics, we are too conservative (Type II error).

We solve the former by leveraging the work on false discovery rate (FDR). As one of the most popular FDR-control procedures, Benjamini–Hochberg [22] controls FDR by comparing p-values from individual hypothesis tests with adjusted thresholds (even under dependency [23]). We can adopt a similar procedure on $L_t(H_1^q)$. Suppose $L_t^{(1)}(H_1^q), \dots, L_t^{(M)}(H_k^q)$ are the $L_t(H_1^q)$ values from each metric sorted in descending order, where $M$ is the total number of metrics. Instead of comparing against a fixed threshold $1/(1 + A_1)$ in Step 1) above, we use

$$L_t^{(m)}(H_1^q) > \frac{1}{1 + \frac{mA_1}{M}}$$

We accept $H_1^q$ when the comparison is true for at least one metric $m = 1, \dots, M$.

On the other hand, to make sure we are not inflating false negatives, we ramp to $q$ when the following two conditions are met: (1) $H_1^q$ is not accepted according to the procedure above, and (2) $H_0^q$ is accepted for the majority of metrics. Here we define majority using the equivalent threshold as $\alpha_0$, i.e. 80%.

## 4.2 Ramping at MPR

Once we have ramped to the Maximum Power Ramp, the criteria to ramp further are quite different from the earlier ramps. The pre-MPR ramps are primarily for risk mitigation, so the SQR ramp framework mainly focuses on trading off speed and risk. On the other hand, since MPR is primarily for measurement, the tradeoff at MPR is mainly on speed vs. decision quality. There are three criteria that the Ramp Recommender considers at MPR, discussed in the following sections.

*4.2.1 MPR Duration.* As we mentioned in Section 3.2, it is important that we spend one week at MPR before making decisions. This is to make sure we capture a representative set of users and use cases throughout a full week cycle, and to take advantage of the reduction of variance as more data trickle in over time. The Ramp Recommender only starts to kick in after the experiment is at MPR for one week.

*4.2.2 Metric Impact.* Clearly, if some metrics are significantly negatively impacted, further ramping beyond MPR is not recommended. Again, the challenge here is to control false positives. As mentioned in Section 4.1.4, there are usually over one hundred metrics that are closely monitored for every experiment. However, not every metric is created equal, especially when it comes to the *final* ramp decision. For a handful of metrics that are most important for the experiment, to keep the recommendation transparent and interpretable, we simply use the same statistical significance definition as shown on our experiment dashboard. If any of these metrics are significantly down (p-value < 0.05), we would like the experimenters to take a closer look instead of recommending further ramp. For the majority of the other metrics, we use false discovery rate to control the number of false alarms. Here we use false discovery rate of 0.1. We use the following steps to determine if any metric impact is significantly negative.
1) Let $p_m$ be the p-value for metric $m$, $(m = 1, ... M)$. We first rank $p_m$'s in increasing order: $p_{(1)}, p_{(2)}, ... p_{(M)}$.
2) Find the largest $l$, such that $p_{(l)} \leq l * \frac{0.1}{M}$
3) If such $l$ exists, and there exists $j, j \leq l$, such that the impact for metric $j$ is negative, we have identified at one metric with significant negative impact. We cannot ramp to 100%. Otherwise, we can ramp to 100%.

Note that if there are 100 metrics, the Ramp Recommender will not recommend ramping to 100% if there are any metrics that are negatively impacted with p-value less than 0.001 (0.1/100).

*4.2.3 Alarming Insights.* If an alarming insight is detected during the ramp, we need to take extra caution when making ramp decisions. These insights include burn-in effect, inconsistent results across ramps, heterogeneous treatment effects etc. Such insights are automatically computed and can be leveraged by the Ramp Recommender to make better, and more informed recommendations.

### 4.3 Evaluation

In this section, we want to evaluate how the Ramp Recommender algorithm performs by replaying it on historical experiments. We primarily focus on evaluating the pre-MPR phase, as we cannot easily quantify decision quality on MPR ramps based on historical data. As we mentioned earlier, during the pre-MPR phase, the key is to tradeoff effectively between speed and risk. There are two aspects we evaluate:

- **Consistency**. How consistent is the recommendation over time? Ideally, if the algorithm recommends ramping up with data collected by time $t$, the same recommendation should hold with data collected by time $t + 1$.
- **Speed**. With ramp recommendation, how much time would we save? Ideally, we should save on both the number of ramps and the total duration before reaching MPR.

We collected 484 experiments ran in the past year, which had one ramp at 50% that lasted for at least a week. These experiments followed various different ramp sequences historically. Therefore, to reply the ramp recommendation, we take the data from the 50% ramps, and simulate the results for any pre-MPR ramp $q \in \{1\%, 5\%, 10\%, 25\%\}$.

Table 1 below compares the recommendation at 5% ramp after Day-1 vs. after Day-7 where all seven days' data are available to make recommendations.

**Table 1: Replayed recommendations show consistency.**

|  | Day-7 Fail | Day-7 Ramp up |
|---|---|---|
| **Day-1 Fail** | 8% | 1% |
| **Day-1 Wait** | 2% | 31% |
| **Day-1 Ramp up** | 0% | 58% |

Note that this replay is for us to evaluate whether the same recommendation holds over time. In reality, we would not have observed Day-7 recommendations for experiments that are ramped up after Day-1. In addition, while there is an option to "wait" for more data on Day-1, the algorithm defaults to ramp up for "undecided" cases after Day-7. As we can see from the table, 58% of these 484 experiments are recommended to ramp up after one day, and all these recommendations would hold even if we had run the experiment for an entire week. Results from other pre-MPR ramp percentages are similar and hence are omitted here.

We have also replayed the entire pre-MPR process for each of the 484 experiments. Given that we do not have prior risk assessment, we choose 1% ramp as the initial ramp to be on the safe side, but use the low risk priors to ramp up from there on. Guided by our algorithm, experiments can either be ramped up to MPR or be terminated halfway. As you can see from Table 2, 71% of the experiments are recommended to ramp to MPR while the rest 29% are deemed too risky to ramp further. These are likely experiments that intended to move metrics. Most experiments are flagged down at 1% ramp and 25% ramp. For those experiments that are recommended to ramp to MPR, we have also compared the duration from the replay with the actual duration of the experiments. Under SQR, it takes 2.9 ramps on average (including the initial 1% ramp) or 12 days in median to reach MPR. This implies about 50% time saved comparing with the actual duration.

**Table 2: Replayed recommendations for entire pre-MPR phase.**

|  | Flagged during Pre-MPR | | | | Reached MPR |
|---|---|---|---|---|---|
| **Last Ramp%** | 1% | 5% | 10% | 25% | 50% |
| **% of Experiments** | 9% | 3% | 4% | 13% | 71% |

## 5 AUTO-RAMP

We have so far discussed the principles and algorithms that determine whether an experiment should ramp up, and if so, to what percent of traffic. In this section, we briefly cover the engineering infrastructure that takes the recommendation and automatically ramps up the experiment accordingly. As the backbone for auto ramping, system infrastructure is designed to achieve the following goals.

- **Reliable**. Every ramp decision could potentially impact thousands if not millions of users, auto-ramp system needs to be able to failover and retry when needed, and all failures and progress need to be closely monitored and communicated to stakeholders.
- **Timely**. Ramp action itself is time sensitive as it correlates to product feature ramping timeline. The data required to make ramp recommendation also need time to accumulate. This requires a robust scheduling system to deliver execution in a timely manner.

Auto-ramp infrastructure is a multi-component system, including an easy to use user interface (UI) for various ramping configurations and user inputs, a middle layer application that interacts with UI and manages metadata, a highly concurrent execution engine that executes ramp recommendation, a distributed and fail safe scheduling system handling time sensitive tasks and monitoring and alerts modules as ramp progresses.

*Setup.* To balance between flexibility and simplicity, the auto-ramp system asks for only critical inputs from users, including risk level assessments, and completion/failure criteria. An auto-ramp is considered completed if it reaches the maximum ramp percentage the experimenter selects, which is not always 100%. For example, some experimenters prefer to exit auto-ramp mode after the experiment reaches MPR. An auto-ramp is considered failed if it passes its preset due date. This can happen when the Ramp Recommender does not have enough information to confidently make the recommendation. These configurations, as part of persisted ramp metadata, are also used for initial ramping plan recommendation.

*Approval.* After the auto-ramp is setup for an experiment, it is then sent to SREs (Site Reliability Engineers) and other key stakeholders for approval, as part of the regular experiment activation process. Once the activators review the setup, validate the risk level and recommended initial plan, and approve the requests, the auto-ramp process is kicked off. Auto-ramp follows the "design - publish - execution" paradigm. Upon approval, its state freezes and becomes read-only for future executions.

*Execution.* Execution is triggered based on frequency, time range and time zone. According to the frequency configured, the execution engine periodically checks if the auto-ramp is overdue or completed. Otherwise, it queries the Ramp Recommender based on the configuration and executes the recommendation.

## 6 SUMMARY AND FUTURE WORK

In this paper, we discussed how we can effectively ramp an experiment by balancing among speed, quality and risk. We first established the SQR framework, offering four principles that practitioners can easily follow. Following these principles, we also developed statistical algorithms that recommend ramp decisions and engineering infrastructure that automatically ramps up the experiment accordingly.

One interesting and related problem we have not studied in depth is deciding when a long-term holdout is beneficial and how to conduct it effectively. Google has shared some long-term experiments they conducted to quantify user-learning effects in the context of Ads [24]. We need a generic solution that can become part of the experimentation process for any experiment.

Another area we can improve upon in our current Ramp Recommender algorithm is on how to decide it is safe to ramp up during the pre-MPR phase in the case of multiple metrics. The ramp criteria we have proposed compromises between risk and speed, but we do not have theoretical guarantees that it controls $\alpha_0$ as we do in the single metric case. From our literature search, this does not seem to be an extensively studied area.